\documentclass[conference,letterpaper]{IEEEtran}

\usepackage[OT1]{fontenc}

\usepackage{amssymb}	 
\usepackage{amsmath}

\usepackage{graphicx}
\usepackage{booktabs}
\usepackage{multirow}
\usepackage{tabularx}
\usepackage{threeparttable}

\usepackage{balance}

\usepackage{url}

\binoppenalty=\maxdimen
\relpenalty=\maxdimen

\DeclareMathOperator{\nn}{\mathbb{N}}
\DeclareMathOperator{\condition}{c}

\DeclareMathOperator{\union}{\cup}

\newcommand{\vc}[1]{\boldsymbol{#1}}

\newcommand{\vcphi}{\vc{\phi}}
\newcommand{\vcvarphi}{\vc{\varphi}}

\begin{document}

\title{Investigating the Agility Bias in DNS Graph Mining}
\author{
\IEEEauthorblockN{Jukka Ruohonen}
\IEEEauthorblockA{Department of Future Technologies \\
University of Turku, Finland \\ 
\texttt{juanruo@utu.fi}}
\and
\IEEEauthorblockN{Ville Lepp\"anen}
\IEEEauthorblockA{Department of Future Technologies \\
University of Turku, Finland \\ 
\texttt{ville.leppanen@utu.fi}
}
}

\maketitle

\begin{abstract}
The concept of agile domain name system (DNS) refers to dynamic and rapidly changing mappings between domain names and their Internet protocol (IP) addresses. This empirical paper evaluates the bias from this kind of agility for DNS-based graph theoretical data mining applications. By building on two conventional metrics for observing malicious DNS agility, the agility bias is observed by comparing bipartite DNS graphs to different subgraphs from which vertices and edges are removed according to two criteria. According to an empirical experiment with two longitudinal DNS datasets, irrespective of the criterion, the agility bias is observed to be severe particularly regarding the effect of outlying domains hosted and delivered via content delivery networks and cloud computing services. With these observations, the paper contributes to the research domains of cyber security and DNS mining. In a larger context of applied graph mining, the paper further elaborates the practical concerns related to the learning of large and dynamic bipartite graphs.
\end{abstract}



\section{Introduction}

The terms DNS agility and agile DNS refer to highly dynamic mappings between fully qualified domain names (FQDNs) and Internet protocol (IP) addresses~\cite{Berger16}. This empirical paper examines the biases from this kind of agility for data mining that operates with DNS graphs comprised of addresses, domains, and links (edges) between addresses and domains.

Content delivery networks (CDNs), as pioneered particularly by Akamai Technologies, Inc., have had a significant impact upon this type of DNS agility. In order to improve reliability and efficiency, these networks essentially distribute the payload both across multiple servers and geographically across multiple locations. In a sense, therefore, CDNs can be viewed as a visible historical milestone in the emergence of cloud computing, the critical juncture having been located in the late 1990s and early 2000s during which Akamai introduced its core commercial technologies. Although the company's infrastructural empire has grown ever \text{since---currently} covering over two hundred thousand servers around the globe~\cite{Akamai15}, the core DNS solution is still based on the same underlying rationale. In a nutshell: a client's DNS query is used to route the client to a topologically nearest and generally optimal caching edge server, provided that the queried domain has joined to a CDN by pointing its name servers to those provided by the CDN orchestrator~\text{\cite{Hablinger11, Holz08, WangShen17}}. While CDNs have been extensively studied in the fields of communications and computer networks, a different kind but still empirically highly similar DNS agility has been at the center of empirical DNS mining applications motivated by cyber security questions.

The malicious use of DNS became widely known by the late 2000s detection of a botnet that generated domain names dynamically. While the botnet used a traditional worm-like propagation to spread, it had a centralized command and control unit to which the bots connected with their daily routines for seeking out the pseudo-random domain names~\cite{Asghari15}. Around the same time, also the so-called fast flux service networks became widely known. While botnets have adopted fast flux algorithms, the phenomenon itself generalizes to CDN-like distributed proxy networks built on top of compromised hosts, which via DNS redirect the traffic to central hosts that serve the actual malicious web or other content~\cite{Holz08}. In this paper, the study of agile DNS is motivated by revisiting two well-known~\cite{Holz08, Zhou15} metrics for observing the behavior of fast flux networks and their hypothesized empirical differences to CDNs and other legitimate DNS-based networking solutions.

The agility bias is operationalized with two criteria that emphasize the statistical effects upon the learning and underlying structure of DNS graphs. In essence, the first question examined in the empirical experiment is simple: (a)~\textit{how many DNS resolving rounds are roughly required to attain a representative DNS graph?} If many rounds are required, it follows that there is a ``learning bias'' for graphs learned with only one or few rounds. In other words, there would be many false negative vertices (absence of vertices that should be present) and false negative edges (missing of relations that should be present) \cite{WangLeskovec12}. The second question examines a different kind of a ``sampling bias'': (b)~\textit{what is the effect of extremely agile domains upon the statistical characteristics of already learned DNS graphs}? While the first question relates to obtaining a statistically representative sample, the sampling bias is best understood as a question about the impact of outliers for the structure of a DNS graph. Here, an outlier is understood broadly as a graph object that is rare and differs considerably from the majority of graph objects sampled \cite{Akoglu15}. Fast flux networks are a good example of such outliers seen in empirical DNS graph mining applications.

To answer to the two questions, the remainder of this paper first briefly revisits the scholarly background in Section~\ref{section: background}, proceeding to discuss operationalization in Section~\ref{section: measurement} as a preparation for the empirical experiment in Section~\ref{section: experimental results}. Conclusions and discussion follow in Section~\ref{section: discussion}.

\section{Background and Related Work}\label{section: background}

The scholarly background behind the concept of DNS agility has been strongly motivated by network security \cite{Berger16}. By and large, the same applies to empirical DNS graph mining approaches. Therefore, the following discussion incrementally proceeds from the fast flux context to the measurement aspects.

\subsection{Flux Types}

The so-called fast flux networks refer to a bundle of largely unknown networking algorithms used to rapidly shuffle domain-address mappings in order to evade detection, improve reliability, and to control a botnet or other malicious network. 




\subsubsection{Domain and Address Fluxing}

It is possible to separate domain name and IP address agility from each other \cite{Hands15}. The former (a.k.a.~domain fluxing) refers to rapid generation of domain names algorithmically. The notorious Conficker worm is likely the earliest known historical example of a network using this technique---the underlying botnet that was constructed via worm-like infections eventually generated even up to fifty thousand domain names each day~\cite{Asghari15, Hands15}. Thereafter, these domain generation algorithms (DGAs) have been successfully used in a number of high-profile botnets, including such technically sophisticated but morbid cases as Rustock~\cite{TheRegister11} and Gameover ZeuS~\cite{Andriesse13}, as well as in related malware implementations such as the CryptoLocker ransomware \cite{Sood16}. Although no DGAs are supposedly present in the empirical experiment, it goes without saying that the potential presence of DGAs may induce a severe bias for DNS graph mining due to the dynamic addition of vertices, including those vertices that are labeled to represent domain names.

Clearly, DGAs are generally only applicable for characterizing malicious agility; no legitimate DNS solution should generate thousands of domain names, and then pick one of these for a brief communication session with a command-and-control unit by activating and subsequently deactivating the associated DNS records~\cite{Demertzis15}. In this paper, however, the interest is to observe the latter type of DNS agility (a.k.a.~IP address fluxing); that is, the dynamic mapping of multiple IP addresses to a single fully qualified domain name. This type of fluxing is common for both legitimate and malicious agility.

\subsubsection{Fluxing by Record Types}

The fast flux phenomenon was historically related to (address) fluxing of IPv4 addresses (a.k.a.~A records). Soon after the discovery of this type of fluxing, malicious networks were observed to also shuffle their name server (NS) records, adding the concept of ``double fluxing'' to the scholarly literature \cite{Zhou15, Hu11, Ruohonen16SNAMS}. A network architecture for double fluxing can be complex. For instance, sometimes the compromised hosts may carry their own malicious DNS servers, which enables a highly customizable network for communication, control, and reliability.

Although single and double fluxing with A and NS records have supposedly remained the mainstream of malicious DNS agility, the versatility of DNS has continued to offer many further possibilities for protocol misuse. To provide further obfuscation guards against detection, DNS tunneling is possible via the text (TXT) records \cite{SANS13, Jin15}. These tunneling techniques provide further covert channels particularly for the command and control aspects. Also AAAA records may be used for address fluxing \cite{SANS16a}, which also motivates the inclusion of IPv6 in the empirical experiment.

\subsection{Quantities}

The seminal fast flux work of Holz and associates~\cite{Holz08} considered two ``unaggregated'' DNS quantities alongside one ``aggregated'' one. The unaggregated and aggregated quantities are, respectively: the number of unique A records (IPv4 addresses), the number of unique NS records, and the number of unique autonomous system numbers for all of the A records. To elaborate the metrics derived from the unaggregated quantities, a couple of clarifications should be made beforehand.

\subsubsection{Aggregation}

In this paper the focus is restricted to the ``unaggregated'' level, but only in the sense that no attempts are made to analyze DNS agility at the level of larger aggregates, including subnets, IP address blocks~\cite{Berger16}, and even Internet registries. The same applies to FQDNs. While ``aggregation'' to the second highest level (cf.~\texttt{xboxlive.com}) is used for one sample in the experiment, which is a common practice in DNS mining \cite{Manadhata14, Ruohonen16MESSA}, the analysis uses also a sample containing ``unaggregated'' domains. For instance, one observed FQDN~is \texttt{download.gfwl.xboxlive.com}, which, through numerous  canonical name records (CNAMEs), is aliased to a content delivery network.

\subsubsection{Time}\label{subsec: time}

It is also important to understand that the study time refers to $i = 1, \ldots, q$ iteratively made DNS queries from a local resolver, using live DNS for the resolving. In other words, the study time operates with a client-side perspective to DNS, using simple polling to obtain the record sets~\cite{Ruohonen16CompSysTech}. Unlike with comparable query engines~\cite{Hu11}, the polling procedure ignores caching and bypasses questions related to time-to-live (TTL) values for the record sets obtained. This said, with small alterations, an equivalent analysis could be carried out also with flow data from DNS servers.


Depending on a bookkeeping solution, each resolving round can be assumed to contain also one or more timestamps in the calendar time~\cite{Ruohonen16CompSysTech}. If $t_i$ is fixed to the integer-valued timestamp at which the $i$:th query completed, the magnitudes in this strictly monotonic time sequence,
\begin{equation}\label{eq: calendar time}
(t_1, \ldots, t_i, \ldots, t_q),
~
t_i \in \nn ,
t_i < t_{i+1}~\textmd{for all}~1 \leq i <  q,
\end{equation}
depend on local infrastructural factors, including bandwidth and the software solution used for resolving. Consequently, the calendar time is generally irregular. Akin to the initial fast flux work~\cite{Holz08}, this paper does not explicitly observe~\eqref{eq: calendar time}, however, and also the $t$-indices are therefore omitted for convenience.

\subsection{Fluxiness and Cumulative Counts}

To measure the dynamics, the \textit{fluxiness} of a domain has been operationalized as the number of all unique A records in a fixed learning period divided by the A records returned by a single DNS lookup of a given domain name \cite{Holz08}. In theory the same operationalization applies to all conventional DNS records returned for a given resource record type, but in this paper the focus is on the conventional A records (IPv4 addresses) and AAAA records (IPv6 addresses). To fix the basic idea as well as the notation, let $R_i$ denote a set of unique DNS records of specific type returned for the $i$:th query:
\begin{equation}\label{eq: R}
R_i \in \lbrace A_i, AAAA_i \rbrace .
\end{equation}

To use one of the private address space networks from RFC 1918 as an example, a couple of queries could return 
\begin{align}\label{eq: A i}
A_i &= \lbrace \texttt{10.0.0.1}, \texttt{10.0.0.254} \rbrace
\quad\textmd{and}
\\ \notag
A_{i+1} &= \lbrace \texttt{10.0.0.1}, \texttt{10.0.0.2}, \texttt{10.0.0.254} \rbrace , 
\end{align}
such that $\vert A_i \vert = 2$ and $\vert A_{i+1} \vert = 3$. This small private example network would be agile in the sense that the fluxiness score $3/2 = 1.5$ for the $i$:th query did not equal unity. If $A_{i+2}$ would subsequently equal $\emptyset$, such that $\vert A_{i+2} \vert = 0$, the $(i + 2)$:th query could have resulted a so-called NXDOMAIN case, that is, the domain would not have resolved to any IPv4 address. As said, timeouts and other errors may be equally likely in practice. 

With this notation, the fluxiness of a domain at the $i$:th query round can be defined with a function

\begin{equation}\label{eq: fluxiness}
\varphi_{R_i} =
f(R_i, R) = 
\begin{cases}
0 & \textmd{if~} R_i = \emptyset, \\
\vert R \vert~/~\vert R_i \vert & \textmd{otherwise}, 
\end{cases}
\end{equation}
where 
\begin{equation}\label{eq: union}
R = R_1 \union R_2 \union \cdots \union R_q 
\end{equation}
refers to all records obtained through $q$ queries. For legitimate agile DNS, the later values in a vector
\begin{equation}\label{eq: fluxiness scores}
\vcvarphi_R = (\varphi_{R_1}, \ldots, \varphi_{R_i}, \varphi_{R_{i+1}} , \ldots, \varphi_{R_q})
\end{equation}
can be hypothesized to converge relatively fast to a fixed scalar, while slower convergence should be expected for malicious fast flux networks~\cite{Holz08}. In other words, it should be generally easier to learn the record sets for legitimate agile DNS compared to malicious agility. Alternatively, the record cumulation 
\begin{equation}\label{eq: cumulative}
\vcphi_R = \left(
\vert R_1 \vert,
\vert R_1 \union R_2 \vert ,
\ldots ,
\vert R_1 \union \cdots \union R_q \vert 
\right)
\end{equation}
should display a slowly decelerating but diverging growth curves for legitimate and malicious fluxing~\cite{Holz08}. Both $\vcvarphi_R$ and $\vcphi_R$ can be also used for probing agile domains in general.

\section{Measurement Framework}\label{section: measurement}

The paper investigates the effect of particularly \eqref{eq: cumulative} upon the learning and sampling of DNS graphs. Before proceeding to elaborate the agility bias, a brief discussion is required for demonstrating the use of the two fast flux metrics in practice, and for introducing the construction of bipartite DNS graphs.

\subsection{Degrees of Agility}\label{subsec: degrees of agility}

\begin{figure}[th!b]
\centering
\includegraphics[width=\linewidth, height=8.0cm]{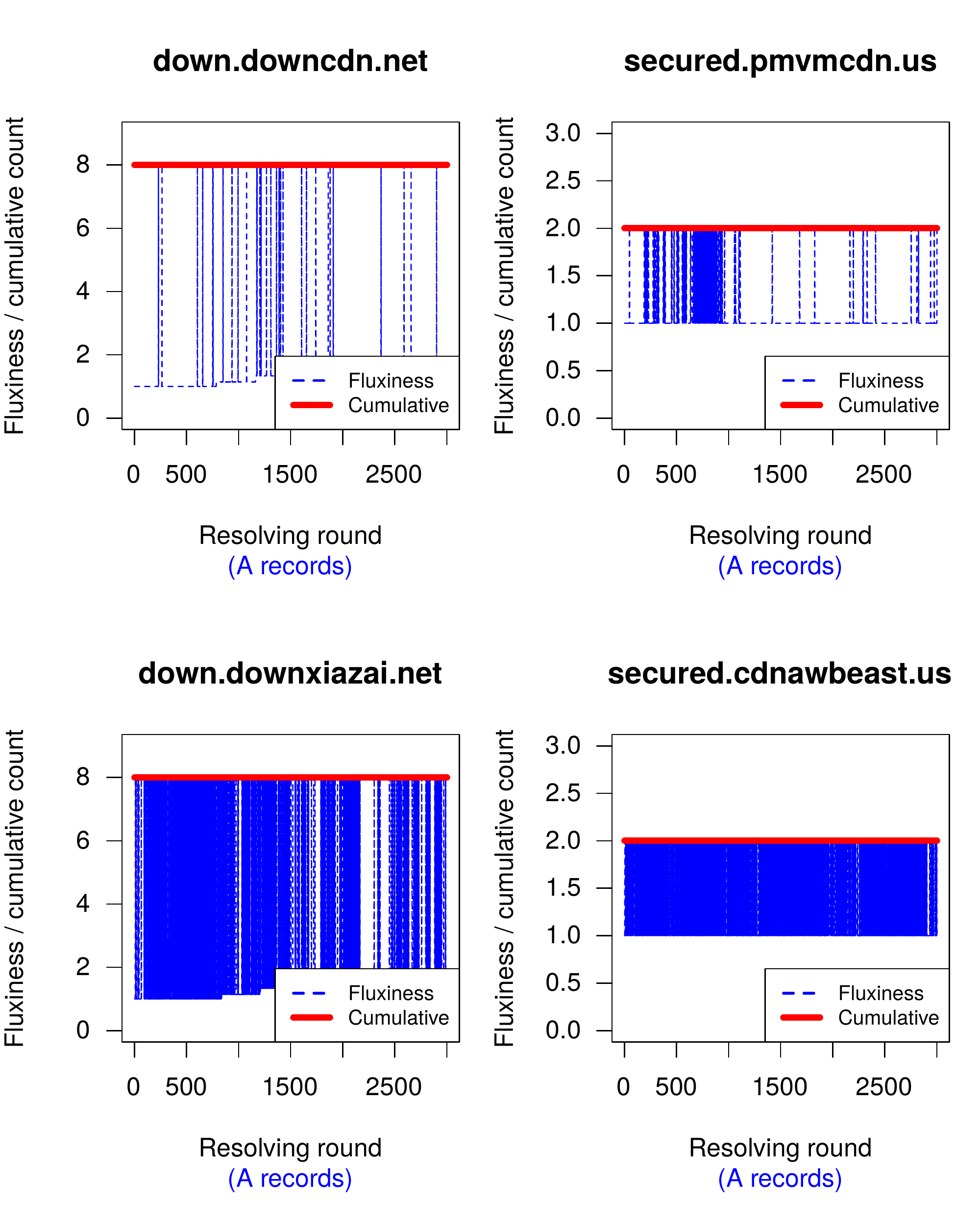}
\caption{A 4-Tuple of Agile Domains (sampled from \cite{malcode})}
\label{fig: flux rapid}
\end{figure}

There exists no single criterion according to which domains could be labeled as agile and non-agile or static and dynamic. When $q \to \infty$, some changes are bound to happen due to infrastructural changes---if only when $t_q$ in  \eqref{eq: calendar time} denotes a sufficiently large integer with respect to $t_1$, such that $t_q - t_1$ amounts to a few years, say. Nevertheless, at minimum, an agile domain should satisfy the following exclusion condition:
\begin{equation}\label{eq: cumulative requirement}
\condition(\tau) : \#~\textmd{of unique values in}~\vcphi_R \geq \tau,~0 < \tau \in\nn,
\end{equation}
for a threshold value $\tau = 2$. In other words, at minimum the size of $R$ must be larger than one for a change to be possible to begin with. This same $\condition(2)$ condition is also equivalent to saying that the standard deviation of  $\vcphi_R$ should be non-zero. To demonstrate the criterion and the unaggregated fast flux metrics in practice, consider the four plots in Fig.~\ref{fig: flux rapid}, which are based on a dataset described later in Section~\ref{subsec: data and sampling}. The solid and dashed lines in the figure refer to \eqref{eq: cumulative} and \eqref{eq: fluxiness scores} for IPv4 addresses, or $\vcphi_A$ and $\vcvarphi_A$, respectively. 


The four visualized domains are clearly agile; there were changes throughout the $q = 3000$ resolving rounds represented on the \text{$x$-axes}. But while all four domains are agile in one sense or another, the degree of agility varies wildly between the domains. The two plots in the top row exhibit rather irregular, stochastic-looking fluxing. The domain \texttt{secured.cdnawbeast.us}, as visualized in the bottom-right plot, exhibits rather extreme agility of different kind by constantly, almost round by round, resolving to either one or two addresses. Despite of these rapid dynamics, however, the cumulative vector $\vcphi_A$ is easy to learn for these domains. In fact, the three thousand resolving rounds were clearly wasteful for learning the domain-address mappings: for all four domains in Fig.~\ref{fig: flux rapid}, the cumulative IPv4 address counts remained constant throughout the resolving rounds.

\begin{figure}[th!b]
\centering
\includegraphics[width=\linewidth, height=8.0cm]{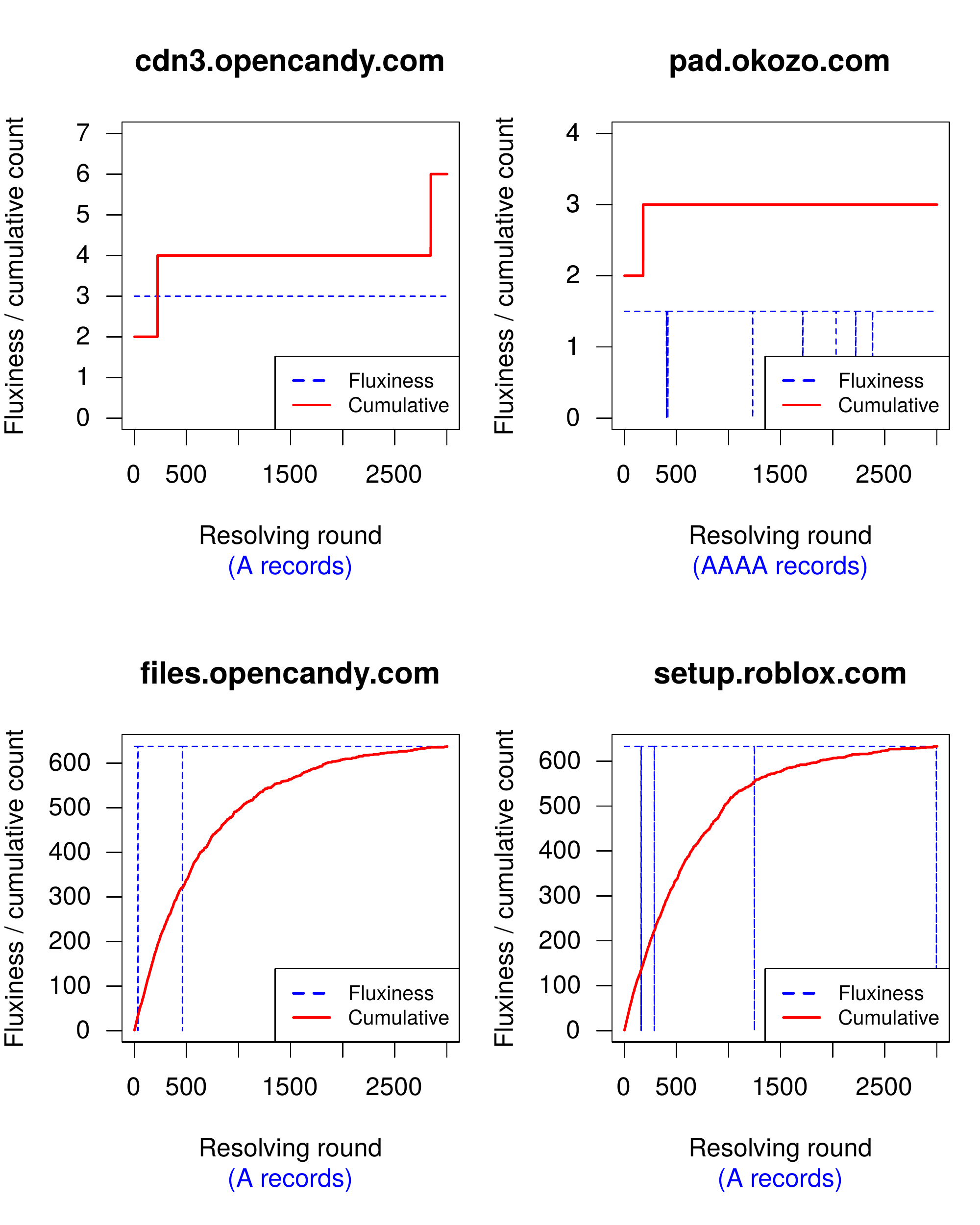}
\caption{Further Agile Domains (sampled from \cite{malcode})}
\label{fig: flux constant}
\end{figure}

Both $\vcvarphi_R$ and $\vcphi_R$ may remain constant around a single unique value. If all records were learned already during the first round, such that $\vert R \vert $ in \eqref{eq: union} equals $\vert R_1 \vert$, the cumulative counts in the vector $\vcphi_R$ in~\eqref{eq: cumulative} would all equal the single unique integer $\vert R_1 \vert$. Likewise, $\vcvarphi_R$ may remain constant. For instance, consider a domain that resolves to two new IPv4s in each $i = 1, 2, 3$ resolving rounds. For such a domain, the function \eqref{eq: fluxiness} results $6/2 = 3$ for each round yet still $\vcphi_A = (2, 4, 6)$. Constant fluxiness scores are also easily demonstrated empirically. An analogous empirical case is thus illustrated in the top-left plot of Fig.~\ref{fig: flux constant} by using the same dataset. This CDN domain \texttt{cdn3.opencandy.com} resolved very slowly to six IPv4 addresses---yet the fluxiness scores remained constant throughout the resolving rounds.\footnote{~According to Wikipedia, ``the OpenCandy ecosystem'' related to adware.} 

When moving to the top-right plot, it becomes evident that also IPv6 addresses may be agile. Moreover, the two bottom row plots both satisfy $\condition(630)$. In other words, more than 21~percent (out of $q = 3000$) of the resolving rounds added a new unique IPv4 address to the cumulative A record sets of \texttt{files.opencandy.com} and \texttt{setup.roblox.com}. Although not that visible from the plots, $\vcphi_A$ still follows a step function also for these two domains. The lengths between the \text{steps---the} number of resolving rounds that resulted the same unique cumulative \text{count---only} widen toward the end of the series, however. It is also worth noting that \texttt{cdn3.opencandy.com} is an alias to \texttt{cdn.opencandy.edgesuite.net}, which is an alias to a domain underneath Akamai's \texttt{akamai.net}. Furthermore, both \texttt{files.opencandy.com} and \texttt{setup.roblox.com} are aliased to Amazon Web Services. Particularly these two cloud-powered domains spotlight important implications for DNS graph mining---clearly, already the shape of the growth curves indicate the presence of a serious bias for ``static'' graphs learned with $q = 1$. In fact, even the $q = 3000$ choice may not have been enough for learning the addresses dynamically mapped to the bottom row domains in Fig.~\ref{fig: flux constant}.

\subsection{DNS Graphs in Brief}

A typical---but by no means unique \cite{Deri13}---DNS graph is undirected and bipartite, connecting fully qualified domain names to their IP addresses~\cite{Berger16, Ruohonen16SNAMS, Kuhrer14}. The basic idea remains similar for server-side applications, although these allow further separating the ``inside'' (origins of queries) and ``outside'' (targets of queries) traffic passing through DNS servers, routers, or related machinery \cite{Jakalan16, Herrmann13, LeeLee14, Yuchi10}. Given the client-side perspective adopted for this paper, the underlying graph structure can be denoted with 
\begin{equation}\label{eq: graph}
G_R = (V_D \cup V_R, E) ,
\end{equation}
where $G_R$ denotes an $R$-type DNS graph given \eqref{eq: R}, $V_D$ and $V_R$ are ordered vertex sets of the observed FQDNs and \text{$R$-type} DNS records, respectively, and $E$ is a set of edges that connect elements in $V_D$ to elements in $V_R$. That is, the edge set $E$ means that there is a relation $E \subseteq \lbrace (v, u) ~\vert~ v \in V_D, u \in V_R \rbrace$ between all $R$-type DNS records $u \in V_R$ to which a domain $v \in V_D$ resolved at~$i$. In addition, a bookkeeping set, say $B_R$, is kept for recording the resolving round at which an $R$-type DNS record was added to the vertex set $V_R$. It follows that $\vert V_R \vert = \vert B_R \vert$ and $b \leq q$ for any positive integer $b \in B_R$. This bookkeeping conveys the dynamic construction of a $G_R$.

Vertices and edges are added dynamically to $G_R$ according to the resolving rounds. That is, for each $i$, new vertices and edges are added to the graph in case these were not already previously added to the graph. To illustrate: for the small IPv4 example (i.e., $R = A$) in~\eqref{eq: A i}, at the $i$:th resolving round there would be a single domain name vertex $v \in V_D$, which would be connected to two IPv4 address vertices, such that $V_A = (\texttt{10.0.0.1}, \texttt{10.0.0.254})$, $(v, \texttt{10.0.0.1}) \in E$, and $(v, \texttt{10.0.0.254}) \in E$. The subsequent round with $A_{i+1}$ would result a slightly larger graph, such that $V_A = (\texttt{10.0.0.1}, \texttt{10.0.0.254}, \texttt{10.0.0.2})$. For this example, $B_A = (i, i, i + 1)$ because the IPv4 addresses \texttt{10.0.0.1} and \texttt{10.0.0.254} were added to $V_A$ during the $i$:th round, while the next round added also \texttt{10.0.0.2}.

This dynamic addition of vertices and edges catalyzes to the learning process in DNS graph mining \cite{Ruohonen16SNAMS}. In a typical data mining application, the vertices and edges are added iteratively either during resolving or afterwards with an order-preserving database collection. When all $i = 1, \ldots, q$ resolving rounds have been processed, the resulting graph can be called a ``learned graph''. Once such a learned graph is available, an observable relational representation is also ready for empirically evaluating the contextual questions of interest.

\begin{figure}[th!b]
\centering
\includegraphics[width=8cm, height=6cm]{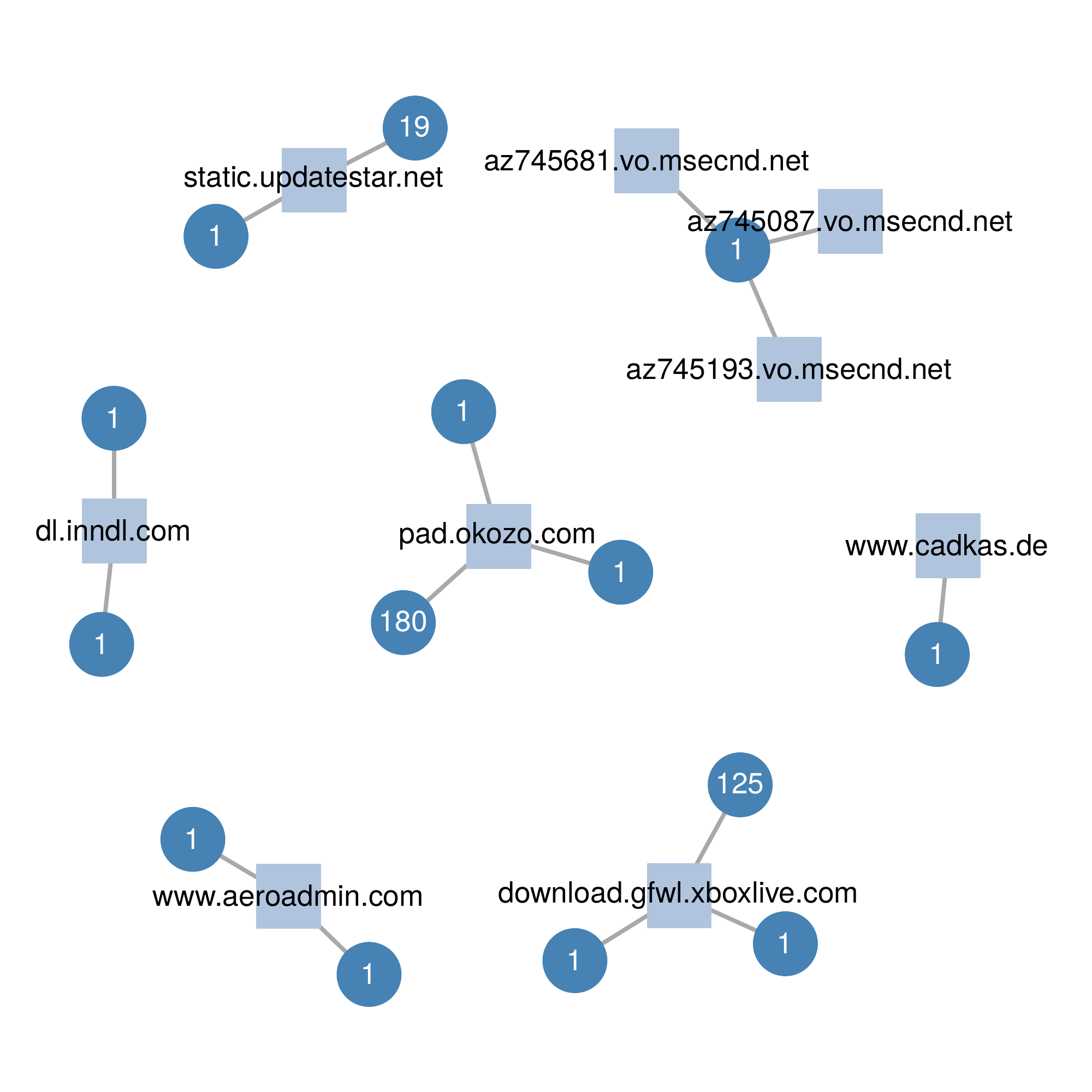}
\caption{A DNS Graph (learned $G_{AAAA}$; sampled from \cite{malcode})}
\label{fig: aaaa}
\end{figure}

Given the ever so slow adoption rate of the IPv6 protocol, an \text{AAAA-based} bipartite DNS graph offers a good small case for further illustrating the bookkeeping structure. Consider thus Fig.~\ref{fig: aaaa}, which shows a learned graph for all IPv6 addresses in the previously noted dataset. Each rectangle denotes a FQDN in $V_D$, while IPv6 addresses in $V_{AAAA}$ are represented by darker colored circles. Instead of using the labels for the AAAA records, each of the circles is labeled with the given $b \in B_{AAAA}$, that is, these labels represent the resolving round at which a given IPv6 address vertex was added to the graph.

As can be seen, the graph is small due to the large amount of domains without active IPv6 addresses. Interestingly, though, many of the shown domains satisfy $\condition(2)$, meaning that the domains eventually resolved to more than one IPv6 address. As an interesting further detail, one AAAA record was also mapped to three domains. More importantly, however, the resolving round integers visualized within the circles indicate that particularly two IPv6 addresses were learned relatively late. From these two domains, the one at the center of the graph is the same \texttt{pad.okozo.com} that was already visualized in the top-right plot in Fig.~\ref{fig: flux constant}. Both figures also convey the same message in terms of the agility bias: if the resolving would have been stopped at $q = 100$, say, the learned $G_{AAAA}$ would have missed two vertices in $V_{AAAA}$ and two edges in~$E$, although $V_D$ would have been unaffected. Thus, the bias would have been rather small. Another way to look at the agility bias relates to the sampling characteristics. If a sample contains extreme cases like the two bottom row domains in Fig.~\ref{fig: flux constant}, the bias is likely a result from outliers rather than overall agility.

\subsection{The Agility Bias}

Agility bias is understood to refer to the statistical effect of the dynamic domain-address mappings upon the construction and underlying quantifiable characteristics of a DNS graph. To study the effects, two different biases are considered next.

\subsubsection{Learning Bias}

The first bias is used to examine the effect of DNS agility on the longitudinal learning of a DNS graph. To carry out the empirical evaluation, an already learned $G_R$ is pruned by iteratively removing vertices and edges in two steps. For each $j = 1, \ldots, q$, (a)~a subset of $R$-type vertices
\begin{equation}\label{eq: learning bias}
S_{R_j} = \lbrace v ~\vert~v \in V_R, b \in B_R, b > j \rbrace, 
\quad S_{R_j} \subseteq V_{R_j},
\end{equation}
is removed from $G_{R_j}$. In terms of the $G_{AAAA}$ graph in Fig.~\ref{fig: aaaa}, the first iteration at $j = 1$ removes $\vert S_{AAAA_1} \vert = 3$ vertices, for instance. To account for the potential disconnected domain name vertices, (b) the graph is pruned afterwards at each step by removing all unconnected vertices with a degree of zero. Thus, the learning bias answers to a ``what if'' question; how much information would have been lost if the learning would have been stopped at some $k < q$? To answer to the question, different graph metrics can be applied at each $j$ to a potentially reduced $G_{R_j}$ that has been processed via the twofold routine.

\subsubsection{Sampling Bias}

The sampling bias relates to a different viewpoint on the biases that agility causes for DNS graph mining. The question asked is: how a learned graph changes when extremely agile domains are excluded from the graph? The question and the term sampling both convey a statistical logic: for instance, if the intention would be to study a sample of non-agile domains resolving to their single unique addresses, even a single case misselection, such as the inclusion of a CDN-powered domain, can cause a severe bias---insofar as the sample should represent a population of non-agile domains.


To measure this kind of a sampling bias, for each domain, the exclusion criteria in \eqref{eq: cumulative requirement} is iterated according to a sequence of threshold scalars $\tau_j \in (1, 2, \ldots, \tau_m)$, $j \leq m$, (a)~removing the corresponding FQDN vertices in $V_D$ for which a given $\condition(\tau_j)$ is not satisfied. It is worth noting that this reduction is exactly the same as removing vertices iteratively according to an ordered vector of degrees for all $v \in V_D$. Analogously to the learning bias, (b) disconnected vertices are afterwards pruned from the processed~$G_{R_j}$. Finally, to fix the iteration, $\tau_m$ is set to the maximum number of unique values in~$\vcphi_R$.  

When considering the agile \texttt{files.opencandy.com} and \texttt{setup.roblox.com} cases in Fig.~\ref{fig: flux constant}, an important question is whether there even is a saturation point for the visualized cumulative counts. While the plots allow to question whether the saturation was reached with $q = 3000$, the answer should still be negative; the counts should not grow without bound~\cite{Zhou15}. That is, the elaborated sampling bias should not be expected to tend to infinity. While speculating about the actual upper bounds is beyond the scope of this paper, it can be noted that the address pools may not be overly large in current CDNs \cite{YinQiao15}. Fast flux networks are a different  beast.

\section{Experimental Results}\label{section: experimental results}

The empirical experiment is carried out by examining the elaborated agility biases for DNS graphs constructed for two separate domain name collections. Before turning to the results, these two FQDN samples should be briefly elaborated.

\subsection{Data and Sampling}\label{subsec: data and sampling}

The first, large sample refers to a single snapshot obtained from the list maintained by OpenDNS (Cisco) for weekly tracking of the most popular domains queried through the open DNS resolvers of OpenDNS \cite{OpenDNS}. The list is updated weekly, each weekly snapshot containing ten thousand domains. The second sample comes from the commonly used (e.g., \cite{LeeLee14}) malc0de domain name collection \cite{malcode}, which is maintained for keeping track of malware files downloaded from the Internet. 

Although the ``benign-versus-malicious'' split should be always done with care \cite{Ruohonen16EISIC}, the two samples can be still characterized to reflect ``popular'' and ``suspicious'' domains. No profiling or other attempts were done to evaluate whether the latter sample would actually contain malicious fast flux networks. As already elaborated, the sample contains content delivery networks and cloud-based domains, nevertheless.

While only one weekly snapshot is used, the OpenDNS sample enlarges to a very large dataset because each of the ten thousand domains were queried for nine hundred times with a custom client-side resolver \cite{Ruohonen16CompSysTech}. As can be seen from the summary in Table~\ref{tab: samples}, also the resolving took a rather long time due to infrastructural and related reasons noted in Section~\ref{subsec: time}. This, said, the calendar time in \eqref{eq: calendar time} is also affected by the ten minute waiting time between resolving rounds in the OpenDNS sample. In contrast, the malc0de sample was resolved with only one minute waiting time.

\begin{table}[th!b]
\begin{center}
\caption{Empirical Samples}
\label{tab: samples}
\begin{tabular}{lccc}
\toprule
& \multicolumn{3}{c}{Sample} \\
\cmidrule{2-4}
& 1.~OpenDNS~\cite{OpenDNS} && 2.~malc0de~\cite{malcode} \\
\hline
Context & Popularity && Malware \\
Resolver & OpenDNS && Local ISP \\
Domains$^a$ & 10,000 && 283 \\
Resolving rounds & 900 && 3000 \\
Time delay$^b$ & 10 && 1 \\
Start of resolving & 29-5-2016 9:35 && 29-6-2016 8:19 \\
End of resolving & ~21-6-2016 10:59 && 5-7-2016 18:45 \\
Aggregation & To the 2nd highest level && No (raw FQDNs) \\
\bottomrule
\end{tabular}
\begin{scriptsize}
\begin{tablenotes}
\item[]{~$^a$~From these, only domains that resolved to one or more A or AAAA records are added to the observed $G_A$ and $G_{AAAA}$ graphs; $^b$~the delay refers to a waiting time (in minutes) between each consecutive resolving round via the reported resolvers.}
\end{tablenotes}
\end{scriptsize}
\end{center}
\end{table}

The OpenDNS and malc0de samples were resolved through live DNS by using the resolvers provided by OpenDNS and a local Internet service provider (ISP). Although a large number of records were obtained, the results are presented by focusing on the ``pure'' DNS graphs comprised of IP addresses and FQDNs. While a domain can obviously have both IPv4 and IPv6 addresses, the analysis is carried out separately for A and AAAA records in order to tentatively evaluate whether the agility biases may generally vary across the two IP versions.

\subsection{Results}

All learned graphs are very sparse, although, interestingly, the learned IPv6 less so (see Table~\ref{tab: descriptive statistics}). The share of domains to IPv4 addresses, $\vert V_D \vert / \vert V_R \vert \times 100$, is about 39~\% and 16~\% for the OpenDNS and malc0de samples, respectively, meaning that both samples are generally comprised of agile domains. Moreover, both $G_{AAAA}$ graphs are rather small, which implies that care should be used when interpreting the IPv6 results.

\begin{table}[th!b]
\begin{center}
\caption{Descriptive Statistics (learned graphs)}
\label{tab: descriptive statistics}
\begin{tabular}{lrcr}
\toprule
& \multicolumn{3}{c}{Sample} \\
\cmidrule{2-4}
& 1.~OpenDNS~\cite{OpenDNS} && 2.~malc0de~\cite{malcode} \\
\hline
Vertices & 29946 && 2063 \\
Edges & 64472 && 1902 \\
Density & 0.0004 && 0.0038 \\
\hline
Vertices & 2911 && 23 \\
Edges & 5859 && 16 \\
Density & 0.0048 && 0.1270 \\
\bottomrule
\end{tabular}
\end{center}
\end{table}

\begin{figure}[th!b]
\centering
\includegraphics[width=8.0cm, height=7.2cm]{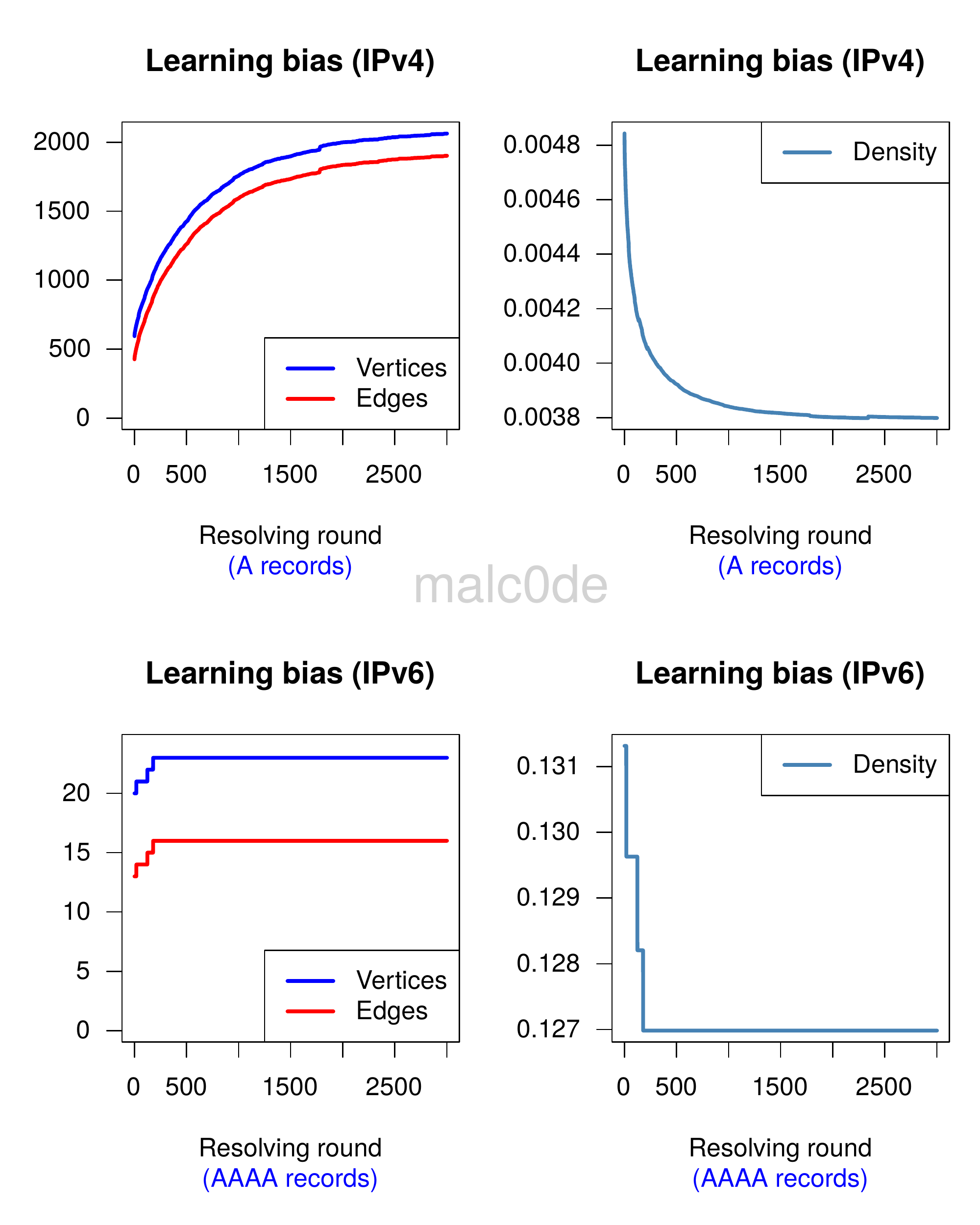}
\caption{Learning Bias for the malc0de Sample}
\label{fig: malcode learning}
\end{figure}

\begin{figure}[th!b]
%
\centering
\includegraphics[width=8.0cm, height=7.2cm]{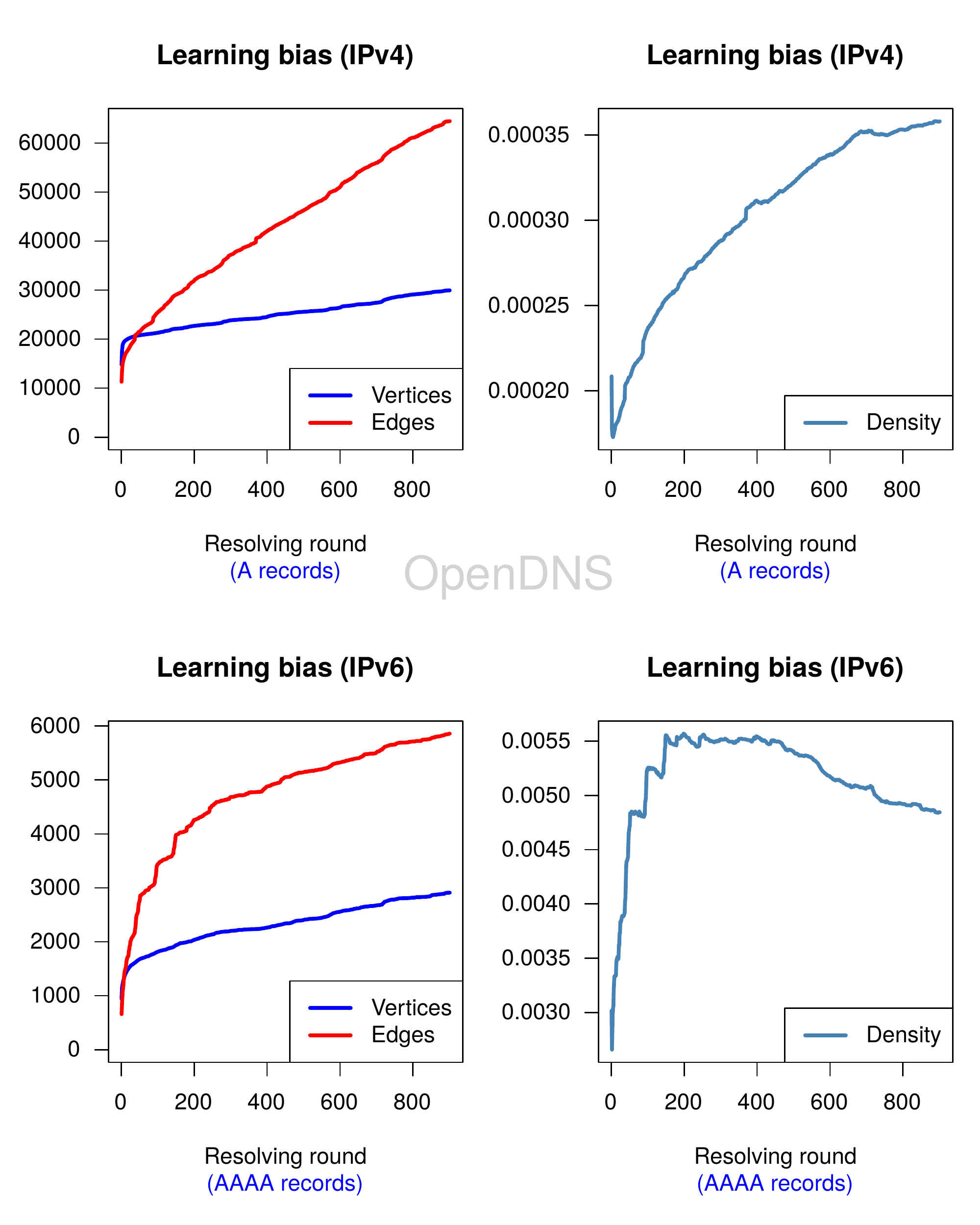}
\caption{Learning Bias for the OpenDNS Sample}
\label{fig: opendns learning}
\end{figure}

The learning biases are shown in Fig.~\ref{fig: malcode learning} and Fig.~\ref{fig: opendns learning}. In both plots, the first rows refers to IPv4 graphs, while results for the AAAA graphs are shown in the bottom rows. The left-hand side plots visualize the number of vertices and edges, whereas the right-hand shows bipartite graph density, $\vert E \vert ~/~ \vert V_D \vert \times \vert V_R \vert$. In all plots, the $y$-axes represent these graph quantities, whereas the $x$-axes denote the resolving rounds.


\begin{figure}[th!b]
\centering
\includegraphics[width=8.0cm, height=7.2cm]{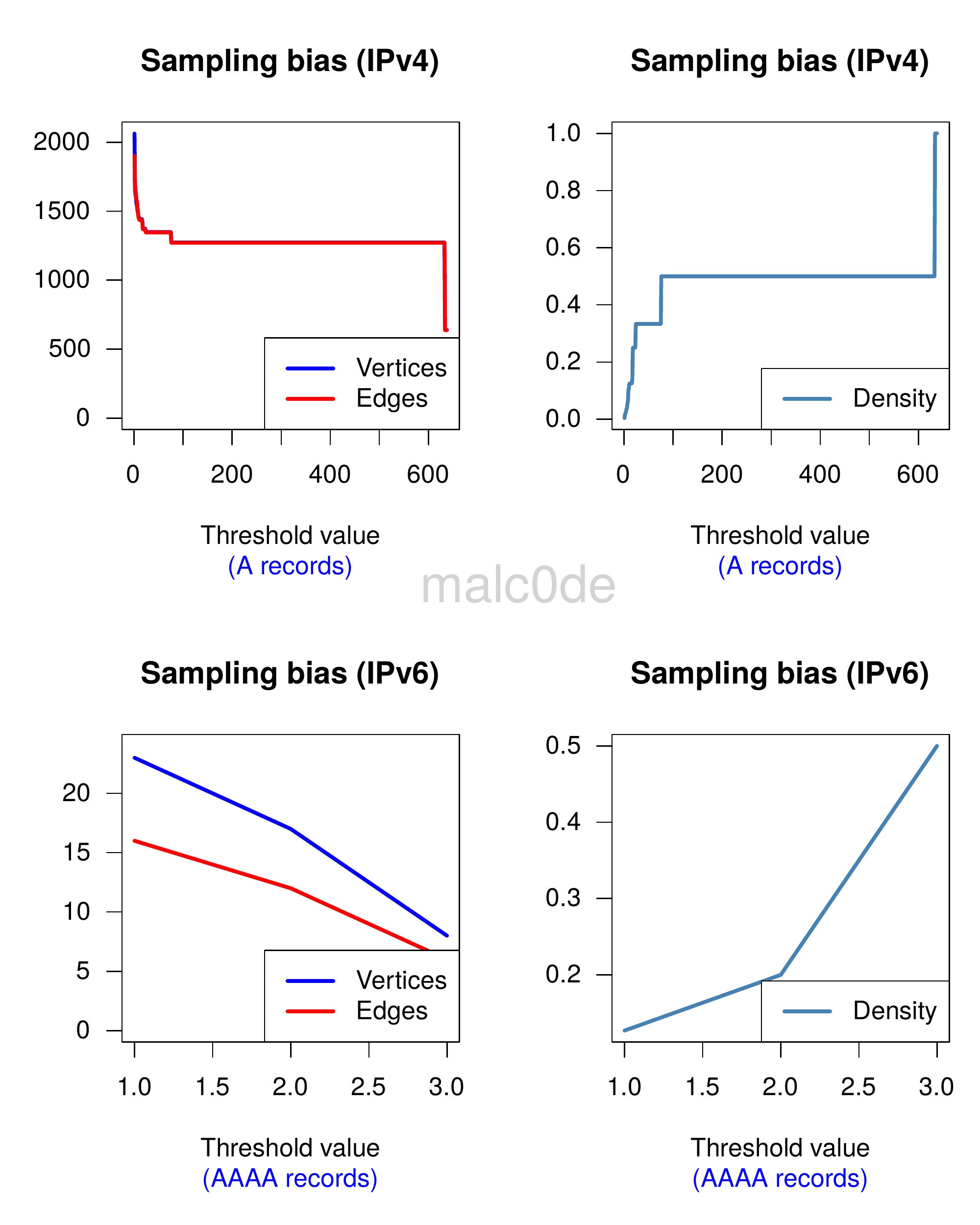}
\caption{Sampling Bias for the malc0de Sample}
\label{fig: malcode sampling}
\end{figure}
\begin{figure}[th!b]
%
\centering
\includegraphics[width=8.0cm, height=7.2cm]{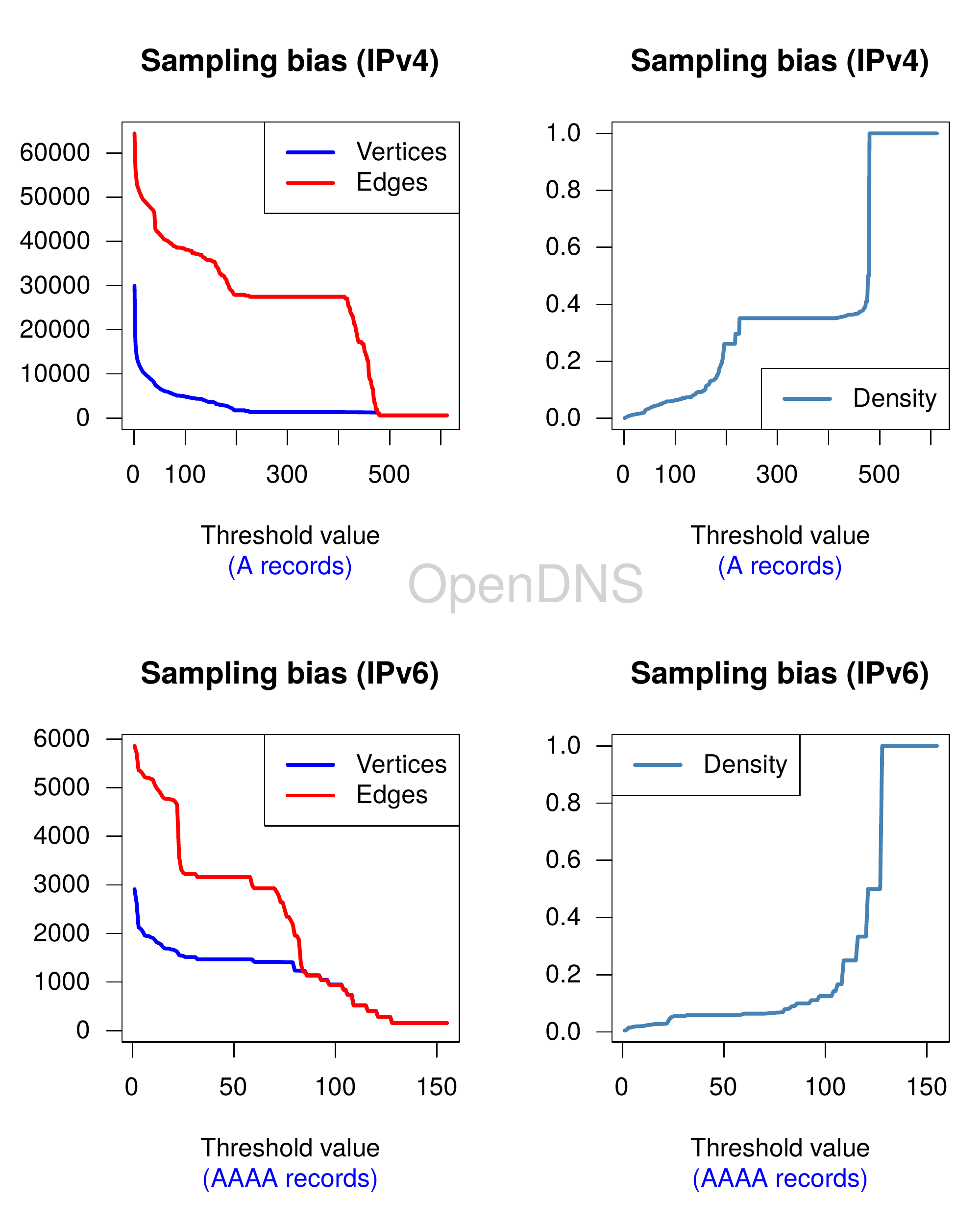}
\caption{Sampling Bias for the OpenDNS Sample}
\label{fig: opendns sampling}
\end{figure}

A brief visual inspection reveals the first observation: while the IPv6 graphs are small for making definite conclusions, the trends are roughly similar between the two IP versions. The learning bias in the malc0de sample follows a highly similar curve than the one shown earlier for the two highly agile bottom-row domains in Fig.~\ref{fig: flux constant}. Due to the general sparsity of DNS graphs, the density for the reduced FQDN-IPv4 graphs declines as learning increases in the malc0de graph, which implies that domains and addresses are constantly added to the graph but in a way that does not connect domain vertices through address vertices. In other words, not many of the domains resolve to IPv4s to which also other domains resolve. 

The contrary applies to the OpenDNS learning bias in Fig.~\ref{fig: opendns learning}. In particular, the top-left plot indicates that learning mostly involves adding new edges in this sample (cf. the red line). Although also new IPv4 addresses are constantly added, the growth rate of the vertex sets is still modest (cf. the blue line). This observation is largely explained by the nature of the sample. Because the dataset contains some of the busiest domains in the Internet, which are mostly served via CDNs and other cloud-based solutions, the IPv4 addresses are learned relatively quickly, and these addresses (or address pools) are used to serve multiple domains in the sample. To summarize, the learning bias can be severe for both samples. If the learning would have been stopped at $q = 100$, for instance, the OpenDNS sample would have missed a considerable amount of edges in particular, and the malc0de sample a considerable amount of both edges and IPv4 vertices.

Turing now to the sampling biases, the top-left plot in Fig.~\ref{fig: malcode sampling} shows that the bias in the malc0de sample is largely a result from the few highly agile, outlying domains. When domains are removed according to the number of their unique A records, the amount of vertices and edges drops sharply relatively early, remaining almost constant thereafter. Therefore, the increasing learning bias in the top-left plot in Fig.~\ref{fig: malcode learning} is largely a result from learning the A records of the outliers, including \texttt{files.opencandy.com} and \texttt{setup.roblox.com} in Fig.~\ref{fig: flux constant}. The sampling biases are less pronounced in the OpenDNS sample (see Fig.~\ref{fig: opendns sampling}), although a similar pattern is present. All in all, also the sampling bias can be severe: a few extremely agile domains can easily dominate the structure of an otherwise relatively stable DNS graph.

\section{Discussion}\label{section: discussion}



This empirical paper examined a so-called agility bias for DNS graph mining with two questions: how many resolving rounds are generally required for constructing domain-address mappings (learning bias), and what is the statistical effect of highly agile domains (sampling bias)? The answers to these two questions is largely in line with the existing recommendation of a two week learning period for typical DNS graph mining applications~\cite{Berger16}. In terms of the pseudo-time resolving rounds, the $q = 3000$ choice used in the paper seems to provide a decent enough benchmark for client-side applications, amounting roughly to two weeks or less. In other words, a rather long period is required even for simple applied questions. To continue the work on mapping of CDN edge server addresses \cite{WangShen17}, a good question for further research would be to examine the potential for an optimal threshold.

When DNS graphs are constructed from server-side flow data in terms of source and destination addresses (see, e.g., \cite{Jakalan16, Fontugne10, Harshaw16}), it should be kept in mind that the destination addresses may be highly dynamic and volatile even though the clients' source addresses may be stable \cite{Manadhata14} or even entirely static. For such applications, the commonly used (see, e.g.,~\cite{Herrmann13, LeeLee14, Yuchi10}) alternative bipartite representations based on client (source) addresses and queried domains (instead of their addresses) may be more robust for measurement. 

All in all, a word of warning can be reasonably reserved for DNS graph mining applications that utilize short learning periods. The severity of the agility bias varies from an application context to another. For instance, malware is often dropped to different file-sharing cloud services~\cite{Ruohonen16EISIC}, which typically utilize either their own agility solutions or outsource the agility to CDNs. Consequently, the agility bias is presumably pronounced in DNS-based malware research, which likely translates to inaccuracies and other issues in machine learning applications for classifying benign and malicious domains.

\section*{Acknowledgments}

The authors gratefully acknowledge Tekes---the Finnish Funding Agency for Innovation, DIMECC Oy, and the Cyber Trust research program for their support.

\balance
\bibliographystyle{IEEEtran}


\end{document}